\documentclass[aps,prl,reprint,superscriptaddress,twocolumn]{revtex4-2}

\usepackage{amsmath,amssymb,mathrsfs,textcomp,esint}
\usepackage{braket} 
\usepackage{graphicx}
\usepackage{dcolumn}
\usepackage{bm}
\usepackage{hyperref}

\begin{document}

\preprint{}

\title{Symmetry-Breaking Electron Dynamics Enable Ultrabroadband Optical-Field Sampling via Second-Harmonic Generation}

\affiliation{Center for Terahertz Waves and College of Precision Instrument and Optoelectronics Engineering, Key Laboratory of Opto-electronics Information and Technical Science, Ministry of Education, Tianjin University, Tianjin 300350, China}
\affiliation{School of Electrical Engineering, Tiangong University, Tianjin 300387, China}
\affiliation{Graduate School, China Academy of Engineering Physics, Beijing 100193, China}
\affiliation{Center for Transformative Science, School of Physical Science and Technology, ShanghaiTech University, Shanghai 201210, China}

\author{Wenqi Tang}
\thanks{These authors contributed equally to this work.}
\affiliation{Center for Terahertz Waves and College of Precision Instrument and Optoelectronics Engineering, Key Laboratory of Opto-electronics Information and Technical Science, Ministry of Education, Tianjin University, Tianjin 300350, China}
\author{Ahai Chen}
\thanks{These authors contributed equally to this work.}
\affiliation{Center for Transformative Science, School of Physical Science and Technology, ShanghaiTech University, Shanghai 201210, China}
\author{Michael Klaiber}
\affiliation{Graduate School, China Academy of Engineering Physics, Beijing 100193, China}
\author{Chunmei Ouyang}
\affiliation{Center for Terahertz Waves and College of Precision Instrument and Optoelectronics Engineering, Key Laboratory of Opto-electronics Information and Technical Science, Ministry of Education, Tianjin University, Tianjin 300350, China}
\author{Yuhai Jiang}
    \email[]{jiangyh3@shanghaitech.edu.cn}
\affiliation{Center for Transformative Science, School of Physical Science and Technology, ShanghaiTech University, Shanghai 201210, China}
\author{Qingzheng Lv}
    \email[]{qzlv@gscaep.ac.cn}
\affiliation{Graduate School, China Academy of Engineering Physics, Beijing 100193, China}
\author{Yizhu Zhang}
    \email[]{zhangyizhu@tju.edu.cn}
\affiliation{School of Electrical Engineering, Tiangong University, Tianjin 300387, China}




\date{\today}

\begin{abstract}

Optical-field sampling using second-harmonic generation (SHG) from strong-field ionization enables ultrabroadband terahertz detection, but the microscopic origin of the SHG signal and its ultrabroadband response have been unclear. Here we show that the target field lifts the half-cycle cancellation of photoelectron dipole emission, generating the SHG signal used for field sampling. Time-dependent Schrödinger-equation simulations, supported by classical-trajectory Monte Carlo analysis, demonstrate that the SHG yield directly encodes the instantaneous target electric field at the ionization time, enabling waveform retrieval by scanning the probe–target delay. Because the SHG response is gated by a subcycle ionization window rather than the probe envelope, the detection bandwidth can extend far beyond the probe duration. We further quantify practical constraints on retrieval, including intrinsic probe asymmetry and SHG back-action, providing a predictive framework to optimize sensitivity, temporal resolution, and fidelity through controlled electron dynamics.

\end{abstract}


\maketitle


In spectroscopy, phase-resolved measurements \cite{kroll2007phase, pandey2024ultrafast} are essential for a complete characterization of optical electric fields. Compared with traditional frequency-domain interferometric techniques \cite{gliserin2022complete, delong1994frequency, iaconis1998spectral} that rely on phase-retrieval algorithms, time-domain sampling methods offer a more direct way to measure the field. In such methods, an ultrashort probe pulse and the target optical field are simultaneously focused into a detection medium, where the medium’s response—such as a current signal—captures the instantaneous value of the target waveform within the probing interval. The duration of the probe pulse is therefore critical for determining the temporal resolution of the measurement. For probe pulses with durations on the order of tens of femtoseconds, applications are typically limited to terahertz waves (0.1–10 THz), as in terahertz time-domain spectroscopy (THz-TDS) \cite{koch2023terahertz,benea2025electro}. Improving temporal resolution naturally suggests compressing the probe pulse width. For example, using the attosecond XUV probe pulse \cite{paul2001observation,vismarra2024isolated,midorikawa2022progress} for photoelectron streaking \cite{goulielmakis2004direct,itatani2002attosecond,carpeggiani2017vectorial,heide2024petahertz, feist2014time, jiang2024time, kiesewetter2018probing} in gases enables sub-femtosecond resolution for detection. However, practical pulse compression encounters fundamental constraints, and the substantial experimental demands become increasingly stringent as the probe duration is reduced.

An alternative strategy for enhancing temporal resolution is to judiciously select both the detection medium and the associated response signal, such that the effective detection window becomes narrower than the duration of the probe pulse itself. For example, in the photoelectron streaking by strong field ionization \cite{PhysRevLett.119.183201,korobenko2020femtosecond,sederberg2020attosecond,zimin2021petahertz,karpowicz2021nanoantenna}, the vast majority of free electrons are created only within a narrow time interval surrounding the probe pulse's maximum. The resulting macroscopic photocurrent (PC) composed of these photoelectrons is then perturbed by the target pulse and consequently encodes the information about the target waveform. This detection scheme has demonstrated a temporal resolution that surpasses the traditional femtosecond limit, achieving 0.1 PHz \cite{korobenko2020femtosecond} or higher \cite{sederberg2020attosecond}.

On the other hand, a series of state-of-the-art THz-TDS techniques \cite{dai2006detection,karpowicz2008coherent,tan2022water} have demonstrated tremendously extended detection bandwidth up to 40 THz, while requiring only conventional femtosecond probe pulses. These schemes utilize second-harmonic generation (SHG) in strong field ionization as the detection signal. Compared with the PC-based methods, it is considerably easier to be detected experimentally. Yet the prevailing physical interpretation of the process based on four-wave mixing in centrosymmetric atoms \cite{dai2006detection} cannot explain the huge enhancement in detection bandwidth. Another description is offered by Brunel radiation theory \cite{ kim2008coherent, oh2014generation, babushkin2017terahertz, brunel1990harmonic}, where SHG arises from the dipole radiation produced by a time-varying plasma photocurrent. While successful as a macroscopic model, Brunel radiation still lacks a microscopic explanation of how the target field's temporal structure is encoded into the SHG spectrum. 

In this Letter, we present a microscopic interpretation of SHG-based sampling methods from the perspective of sub-cycle electron dynamics, using the strong field high-order harmonic generation theory. We show that the SHG signal originates from the asymmetry of interfering dipole radiation emitted by electrons ionized in two adjacent positive and negative half-cycles of the probe pulse. The target field modifies both the ionization probability and the subsequent evolution of the electrons from the two half-cycles in a different manner, thereby breaking the inherent symmetry of the interference. By analyzing the target-field-induced variation in the SHG spectrum using both the time-dependent Schrödinger equation (TDSE) and the classical trajectory Monte Carlo (CTMC) \cite{liu2013classical, gan2024probing, zhang2021electron, gao2023coulomb} approach, we accurately reconstruct the waveform of the target field. 
Furthermore, our CTMC analysis demonstrates that the target-field-induced perturbation of the ionization rate is the primary driver of symmetry breaking, significantly outweighing the effect of trajectory modifications in the continuum. These insights rationalize the bandwidth enhancement observed in SHG-based THz-TDS and provide a robust roadmap for optimizing future optical sampling performance.

\begin{figure}
    \includegraphics[width=1\linewidth]{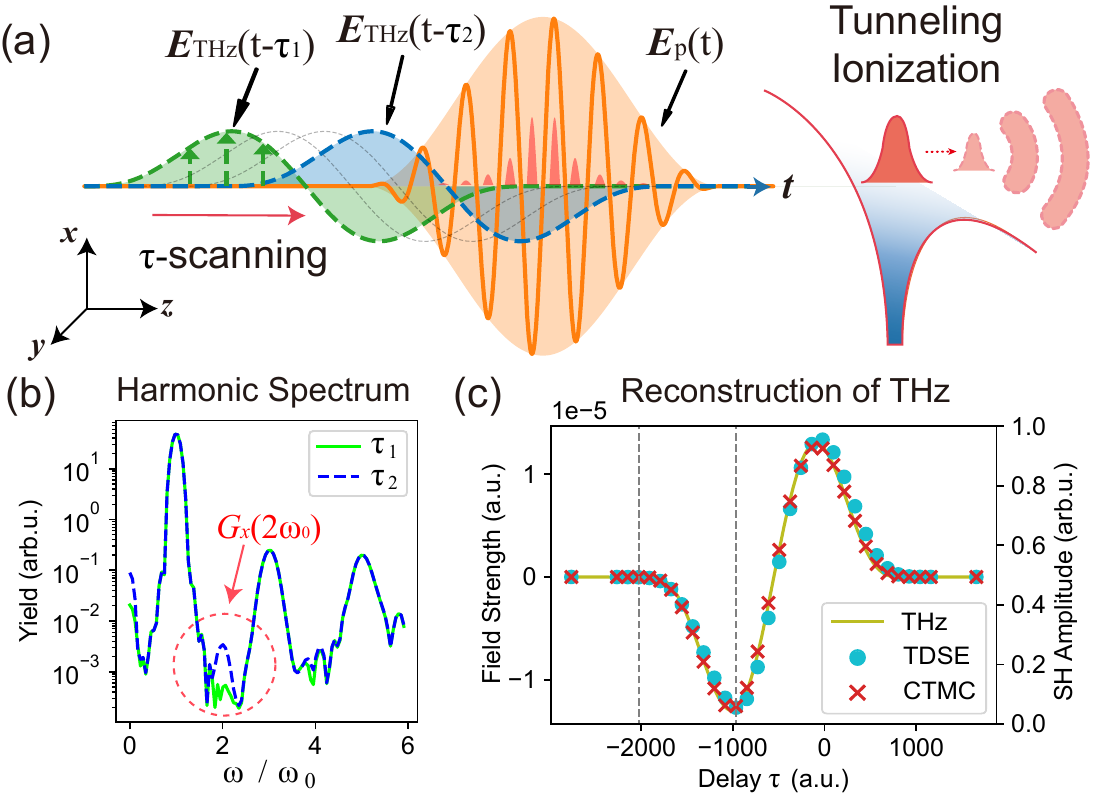}
    \caption{\label{fig:1}(a) Sketch of the system: a strong probe pulse $\boldsymbol{E}_\mathrm{p}$ (orange) and a target field $\boldsymbol{E}_\mathrm{THz}$ (green and blue, for different time delay $\tau_1$ and $\tau_2$, respectively) interact with a hydrogen atom. The THz wave is accompanied by an optional DC bias to achieve the coherent detection in the SHG scheme \cite{dai2006detection, karpowicz2008coherent}, with peak field strength $\mathcal{E}_{\mathrm{THz}}=1.7\times10^{-5}$ a.u. (87 $\mathrm{kV / cm}$), $\mathcal{E}_\mathrm{DC}=2.0\times10^{-5}$ a.u. and central frequency $\omega_\mathrm{THz}=0.04\omega_{0}$ (15 THz). (b) Detectable SHG spectrum and 
    (c) the reconstruction of the THz waveform by the SHG sampling process. The SHG signal form the TDSE simulation (blue dots) and the CTMC prediction (red crosses) are both normalized to the maximum of the THz field strength.
    }
\end{figure}

We model the sampling procedure in Fig.\ref{fig:1}, where a strong probe pulse $\boldsymbol{E}_{\mathrm{p}}(t)$ and an arbitrary target pulse $\boldsymbol{E}_{\mathrm{THz}}(t)$ to be detected are temporally overlapped and interact with a hydrogen atom, triggering strong-field harmonic generation. We assume that the probe pulse is a femtosecond pulse with a central wavelength of 800 nm and peak intensity around $10^{14}$ W/cm$^2$ (Keldysh parameter $\gamma \approx 1$ \cite{popruzhenko2014keldysh}), in accordance with parameters primarily used in experiments \cite{dai2006detection}. The total electric field writes $\boldsymbol{E}(t; \tau)=\boldsymbol{E}_{\mathrm{p}}(t)+\boldsymbol{E}_{\mathrm{THz}}(t-\tau)$, where $\tau$ is the time delay between the two pulses. The probe pulse $\boldsymbol{E}_{\mathrm{p}}(t) = \frac{1}{\sqrt{\epsilon^2+1}}\mathcal{E}_\mathrm{p}f(t)[ \cos(\omega_0 t+\phi)\boldsymbol{\mathbf{e}}_x + \epsilon \sin(\omega_0 t+\phi)\boldsymbol{\mathbf{e}}_y] \quad (0 \leq t \leq T_\mathrm{p}=2n_c\pi /{\omega_0})$, with envelope function $f(t)=\sin^2(\frac{\omega_0 t}{2n_c})$. The number of cycles $n_c=15$, central frequency $\omega_0 = 0.056$ a.u., peak electric field strength $\mathcal{E}_{p}=0.0534$ a.u. and a small ellipticity $\epsilon=0.2$ to suppress the recollision \cite{burnett1995ellipticity,silaev2022using}.
Meanwhile, the target field is a typical one-cycle THz burst $\boldsymbol{E}_{\mathrm{THz}}(t)=\mathcal{E}_\mathrm{THz}\sin^2(\omega_\mathrm{THz} t /2)\sin(\omega_\mathrm{THz}t)\mathbf{e}_x$ linearly polarizing in \textit{x}-direction, corresponding to the experiments. Henceforth, we adopt atomic units (a.u.), unless stated otherwise.

In TDSE simulation, once the evolution of electron's wavefunction $\ket{\Psi(t)}$ is obtained, the spectral distribution of the dipole radiation from the electron can be calculated from \cite{sundaram1990high}
\begin{equation}
    \begin{aligned}
        \frac{\mathrm d \mathcal{N}(\omega)}{\mathrm d \omega} &= \frac{\omega^3}{(2\pi)^2 c^3} \left|\mathbf{e}_\sigma \cdot \int_{t_i}^{t_f} \mathrm d t^\prime \braket{\hat{\boldsymbol{d}}(t^\prime)} e^{i\omega t^\prime} \right|^2 \\
        &= \frac{\omega^3}{(2\pi)^2 c^3} \left|\mathbf{e}_\sigma \cdot \braket{\hat{\boldsymbol{d}}(\omega)} \right|^2 \propto \left| G_\sigma(\omega) \right|^2,
    \end{aligned}
    \label{e1}
\end{equation}
where $\mathbf{e}_\sigma$ denotes the polarization direction of the radiation field, $\braket{\hat{\boldsymbol{d}}(\omega)}$ is the Fourier transform of the dipole moment $\braket{\hat{\boldsymbol{d}}(t)}=\bra{\Psi(t)}(-\hat{\boldsymbol{r}})\ket{\Psi(t)}$, and $G_\sigma(\omega)$ is the phase-involved spectrum. Since the major polarization of probe pulse is along \textit{x}-axis, we let $\mathbf{e}_\sigma=\mathbf{e}_x$.
In Fig.\ref{fig:1}(b), the TDSE results demonstrate that the overlapped target field ($\tau_2$, blue curve) induces a pronounced peak at second harmonic. By scanning the time delay $\tau$, the SH intensity $|G_{\mathrm{SH},\sigma}(\tau)|^2$ contains information of the target field. Fig.\ref{fig:1}(c) indicates that $|G_{\mathrm{SH},\sigma}(\tau)|$ shows excellent convergence with respect to the target THz waveform $\boldsymbol{E}_\mathrm{THz}(t)$, reproducing the sampling process theoretically.

\begin{figure}
    \includegraphics[width=1\linewidth]{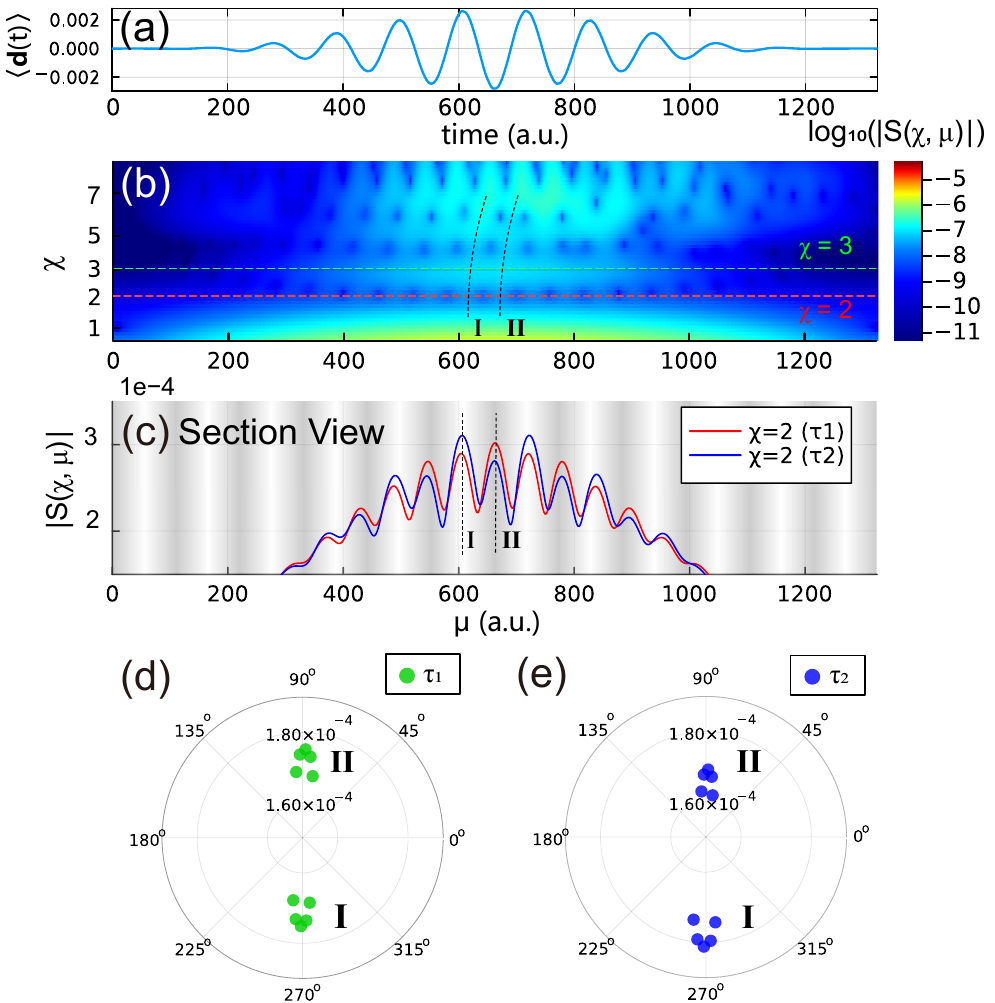}
    \caption{\label{fig:2} 
    (a) The dipole moment $\braket{\hat{\boldsymbol{d}}(t)}$ obtained by TDSE simulation.
    (b) Wavelet transform of $\braket{\hat{\boldsymbol{d}}(t)}$, denoted as $S(\chi, \mu)$ for the time delay $\tau_1$. Labels I and II denote two distinct classes of ionization events occurring in opposite half-cycles of $\boldsymbol{E}_\mathrm{p}$. The vertical dashed lines indicate the contributions from the two most probable trajectories within each respective class. (c) Section views of $|S(\chi, \mu)|$ for $\chi=2$ at time delays $\tau_1$ (red) and $\tau_2$ (blue), with the white-grey gradient background indicating the phase of $S(\chi=2, \mu)$ from 0 to $2\pi$. An obvious asymmetry between I and II appears for the time delay $\tau_2$. (d) and (e) show the contribution of the representative trajectories to the SHG signal in Eq.(\ref{e3}). The trajectories are around maximum ionization rate in I and II. The radius denotes the amplitude while the polar angle is the phase. The imbalance cancellation between I and II happens around $\tau_2$. The other parameters are the same as in Fig.\ref{fig:2}.}
\end{figure}

To understand how does SHG encode the information of target waveform, we first study the formation mechanism of the SHG. The ionized wavepackets and its subsequent oscillations in the total electromagnetic field cause the emission of a harmonic spectrum, spanning from the quasi-static limit to multiples of the fundamental frequency. To capture this dynamics, a time-frequency analysis of $\braket{\hat{\boldsymbol{d}}(t)}$ is performed, as shown in Fig.\ref{fig:2}(a) and (b). We utilize the continuous wavelet transform \cite{guo2022review} with generalized morse wavelet with parameters $\Gamma=5$ and $\beta=1$, so that the instantaneous frequency components confined in a narrow time window can be extracted (see in \cite{Supp}). Mathematically speaking, the wavelet transform of $\braket{\hat{\boldsymbol{d}}(t)}$ is defined as 
\begin{equation}
    \begin{aligned}
        S(\chi, \mu)=\frac{1}{\sqrt{|\chi|}} \int_{-\infty}^{+\infty} \braket{\hat{\boldsymbol{d}}(t)} \mathcal{M}^{*}_{\Gamma=5,\beta=1}\left(\frac{t-\mu}{\chi}\right) \mathrm{d} t,
    \end{aligned}
    \label{e2}
\end{equation}
where $\mathcal{M}(t)$ is the morse wavelet function, which can be regarded as a Gaussian envelop modulated by a complex-value carrier wave with the same frequency as the probe pulse ($\omega_{\psi}=\omega_0$). $\chi$ (proportional to the harmonic order) and $\mu$ (proportional to the emission time for the photonelectrons) are the scale and translation parameters respectively. This approach can be used to extract different frequency components at different time.

The TDSE result of $|S(\chi, \mu)|$ with respect to the time delay $\tau_1$ [green curve in Fig.\ref{fig:1}(a)] is shown in Fig.\ref{fig:2}(b). First of all, there is an obvious difference between low-order harmonics and high-order harmonics. The high-order harmonics exhibits a significant \textit{lag} effect compared to the ionization time, as the two vertical dashed lines bend to the right with increasing $\chi$. This is because the HHG is caused by the re-collision process where the photoelectrons has to gain energy in the laser field after ionization. In contrast, the low-order harmonics are almost synchronized with the ionization, as it originates from the transient change of dipole moment caused by the tunneling wavepacket.

Secondly, the distribution of $|S(\chi, \mu)|$ near $\chi=2$ exhibits pronounced oscillations (red dotted line), whereas it remains smooth for $\chi=3$ (green dashed line). This oscillatory behavior originates from the phase difference between the radiation within the positive and negative half-cycles of the probe pulse. In the absence of the target field (time delay $\tau_1$), the distribution for $\chi$=2 is perfectly symmetric, with each peak differing in phase by $\pi$. Under these conditions, the SHG components from I and II interfere destructively, leading to the well-known suppression of even harmonics in centrosymmetric atomic systems. When the target field $\boldsymbol{E}_\mathrm{THz}$ is superimposed (time delay $\tau_2$), this symmetry between I and II is slightly perturbed [blue curve in Fig.\ref{fig:2}(c)], producing a mismatch in the interference. As a consequence, the destructive interference becomes incomplete, resulting in the appearance of a finite SHG component.


Naturally, a crucial question arises: how does the target field influence the SHG contribution from each electron? More specifically, is the dominant effect due to the perturbation of the ionization rate $w^{(i)}$ (hereafter referred to as the A-factor), or to the modification of the subsequent electron dynamics (the B-factor)? The TDSE simulations struggle to provide an intuitive picture of the underlying mechanism, and therefore, an analysis based on the CTMC method will be useful \cite{liu2013classical, gan2024probing, zhang2021electron, gao2023coulomb}. In CTMC calculations, the SHG could be regarded as the coherent superposition of classical dipole radiation from each trajectory, written as
\begin{equation}
    \begin{aligned}
        \mathcal{G}^{}_{\sigma}(2\omega_0) &= \sum_i w^{(i)} \mathscr{F} \{ \ddot{\boldsymbol{r}}^{(i)}(t) \cdot \mathbf{e}_\sigma \} (2\omega_0) \\
        &= \sum_i w^{(i)} \mathcal{G}^{(i)}_{\sigma} (2\omega_0),
    \end{aligned}
    \label{e3}
\end{equation}
where the \textit{i}-th trajectory is ionized at time $t^{(i)}_\mathrm{ion}$ with the weight $w^{(i)}$ \cite{Supp}. Even though this model neglects the contribution of bound states to the dipole moment, the CTMC calculation can still reproduce the waveform of the target field accurately, as shown in Fig.\ref{fig:2}(a). 

The symmetry breaking between the two classes of ionization events (I and II), induced by the target field, can be evaluated using the CTMC approach. For clarity, we focus on several representative trajectories that are ionized near the positive and negative maximum of the electric-field, and examine their respective SHG contributions—both in amplitude and phase. For time delay $\tau_1$, the SHG contributions from events I and II exhibit nearly complete mutual cancellation [Fig.\ref{fig:2}(d)], reflecting the intrinsic symmetry of the unperturbed system. In contrast, at $\tau_2$, the target-field-induced symmetry breaking is evident, leading to incomplete cancellation [Fig.\ref{fig:2}(e)]. This behavior is fully consistent with the spectral features observed in Fig.\ref{fig:2}(c).

\begin{figure}
    \includegraphics[width=1\linewidth]{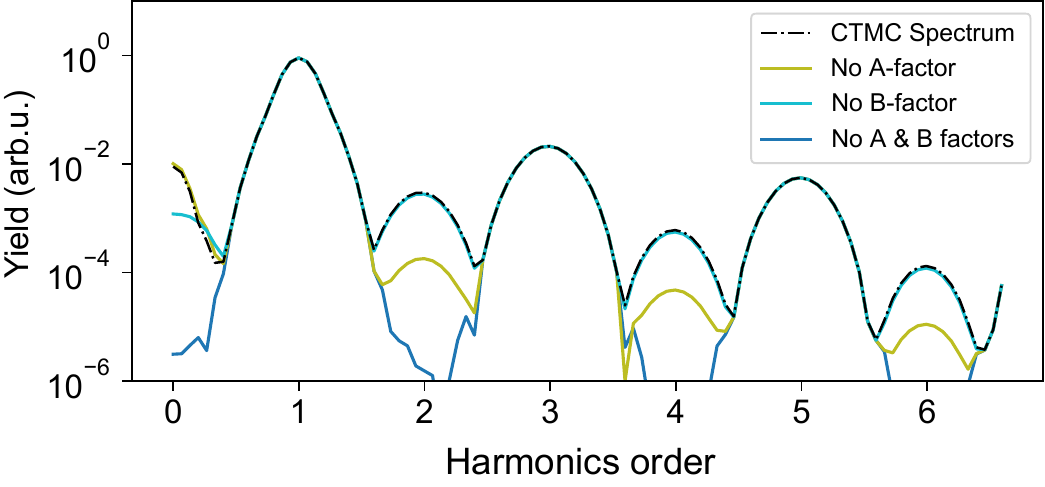}
    \caption{\label{fig:3}Harmonic spectrum $\mathcal{G}_{x}(\omega)$ obtained from CTMC. Black dash dot: No factors are removed; Yellow line: Perturbation of ionization rate (factor A) is removed;  Cyan line: Perturbation of the subsequent motions (factor B) is removed; and Blue line: Both factor A and B are removed. The other paramters are the same as in Fig.\ref{fig:2}.}
\end{figure}

Moreover, the CTMC model offers a natural means to disentangle the two contributions (A-factor v.s. B-factor) by selectively suppressing one factor while keeping the other unchanged. In Fig.\ref{fig:3}, the simulations reveal that the perturbation to the ionization rate plays the primary role in SHG generation. This finding is counterintuitive, given that the target field is several orders of magnitude weaker than the probe field. To validate this conclusion, we examine the interference between two representative trajectories in I and II, and their combined SHG contribution can be approximated as
\begin{equation}
    \begin{aligned}
        \mathcal{G}_{x}^\mathrm{I,II}(2\omega_0) &= \sum_{i \in \{\mathrm{I}, \mathrm{II}\}} w^{(i)} \mathcal{G}^{(i)}_{x} (2\omega_0) \\
        &= \sum_{i \in \{\mathrm{I}, \mathrm{II}\}} w^{(i)} \mathscr{F} \{\mathbf{e}_x \cdot \boldsymbol{E}(t, \tau) \Theta(t-t_\mathrm{ion}^{(i)}) \}(2\omega_0).
    \end{aligned}
    \label{e4}
\end{equation}
As the target pulse $\boldsymbol{E}_{\mathrm{THz}}(t)$ varies slowly within the ionization gate, and the small variation of ionization rate is parameterized as $w^\mathrm{(I)}=w^\mathrm{0}(1+\delta w)$ and $ w^\mathrm{(II)}=w^{0}(1-\delta w)$, Eq.(\ref{e4}) can be eventually approximated as \cite{Supp}
\begin{equation}
    \mathcal{G}^{\mathrm{I,II}}_x(2\omega_0)=w^0\frac{(4/3)\mathcal{E}_\mathrm{p}\delta w + \mathcal{E}_\mathrm{THz}}{i\sqrt{2\pi}\omega_0}.
    \label{e5}
\end{equation}
This compact formula concisely describes the general relationship between the SHG and the two contributions, i.e., A-factor ($\delta w$) and B-factor ($\mathcal{E}_\mathrm{THz}$). According to the ADK rate $W(E)=(4/E)\exp(-2/3E)$, if we apply $\mathcal{E}_{\mathrm{p}}=0.05$ a.u. and $\mathcal{E}_{\mathrm{THz}}=1.7 \times 10^{-5}$ a.u., then the variation in ionization weight $\delta w=W(\mathcal{E}_{\mathrm{p}} + \mathcal{E}_{\mathrm{THz}}) / W(\mathcal{E}_{\mathrm{p}})-1=0.0042$, in which $\mathcal{E}_{\mathrm{p}}\delta w=2.1\times10^{-4}$ substantially exceeding $\mathcal{E}_{\mathrm{THz}}$ in Eq.(\ref{e5}). It shows that the ionization process ensures the prominence of the target field even being several orders of magnitude weaker than the probe pulse.

\begin{figure}
    \includegraphics[width=1.0\linewidth]{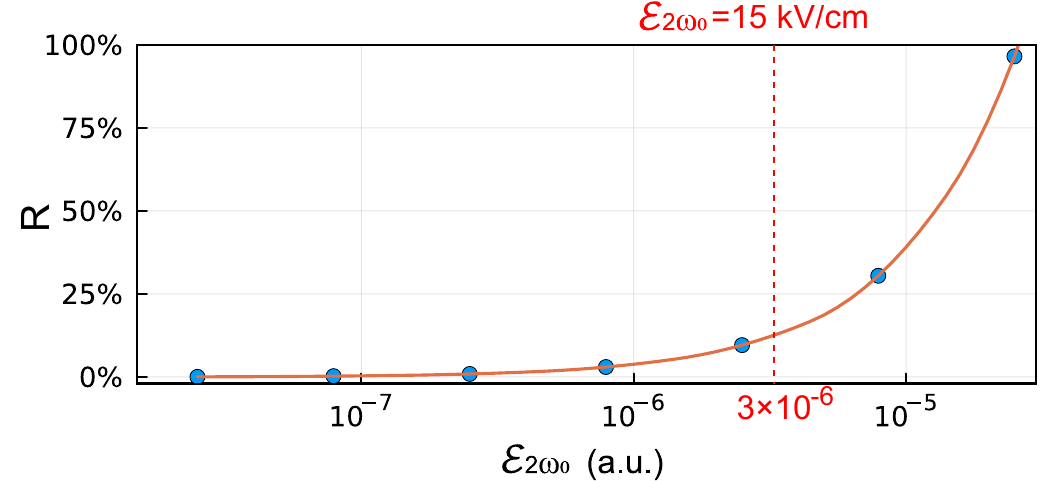}
    \caption{\label{fig:4} The change ratio $R$ as a function of the $2\omega_0$-field strength $\mathcal{E}_{2\omega_0}$. The dashed vertical line corresponds to the field strength reported in the experiment \cite{fujii2021electric,garriga2022observation}. The other parameters are the same as in Fig.\ref{fig:2}. 
    }
\end{figure}

Since a weak field can achieve symmetry breaking and trigger detectable SHG signal, a fundamental question in THz-field-induced SHG is whether the generated second harmonic introduces an additional asymmetry and exerts a back-reaction on the signal itself.
Experimental studies \cite{fujii2021electric,garriga2022observation} reveal that the generated SHG field can reach 15 kV/cm, which is only one order of magnitude below typical THz fields. This suggests that the self-generated second harmonic field is non-negligible. To quantify this back-action, we model SHG driven by the target THz field superimposed with a weak auxiliary $2\omega_0$-field $\boldsymbol{E}_{2\omega_0}(t)=\mathbf{e}_x\mathcal{E}_{2\omega_0}f_{2\omega_0}(t)\cos(2\omega_0 t+\phi_{2\omega_0})$. We check the net SHG $\Delta G_x(2\omega_0)=G^{(\omega_0,2\omega_0)}_x(2\omega_0)-G^{(2\omega_0)}_x(2\omega_0)$, where the contribution from the $2\omega_0$-field alone $G^{(2\omega_0)}_x(\omega)$ is subtracted \cite{Supp}. The change ratio between the net SHG amplitude and the pure THz-induced SHG $G^{(\omega_0)}_x(2\omega_0)$ defined as $R = |\Delta G_x(2\omega_0) - G^{(\omega_0)}_x(2\omega_0)|/|G^{(\omega_0)}_x(2\omega_0)|$ is shown in Fig.\ref{fig:4}. It reveals that the $2\omega_0$-field enhances the net SHG by around 10\% at a field strength of $\mathcal{E}_{2\omega_0}=3\times10^{-6}$ a.u. (15 kV/cm). The net SHG grows nonlinearly with $\mathcal{E}_{2\omega_0}$, implying that the accumulation of the emitted second harmonic may significantly alter the total SHG yield. These findings offer a compelling mechanism for zero-DC bias coherent detection \cite{dai2006detection} and highlight two-color field engineering as a potential degree of freedom for controlling SHG dynamics.

\begin{figure}
    \includegraphics[width=1\linewidth]{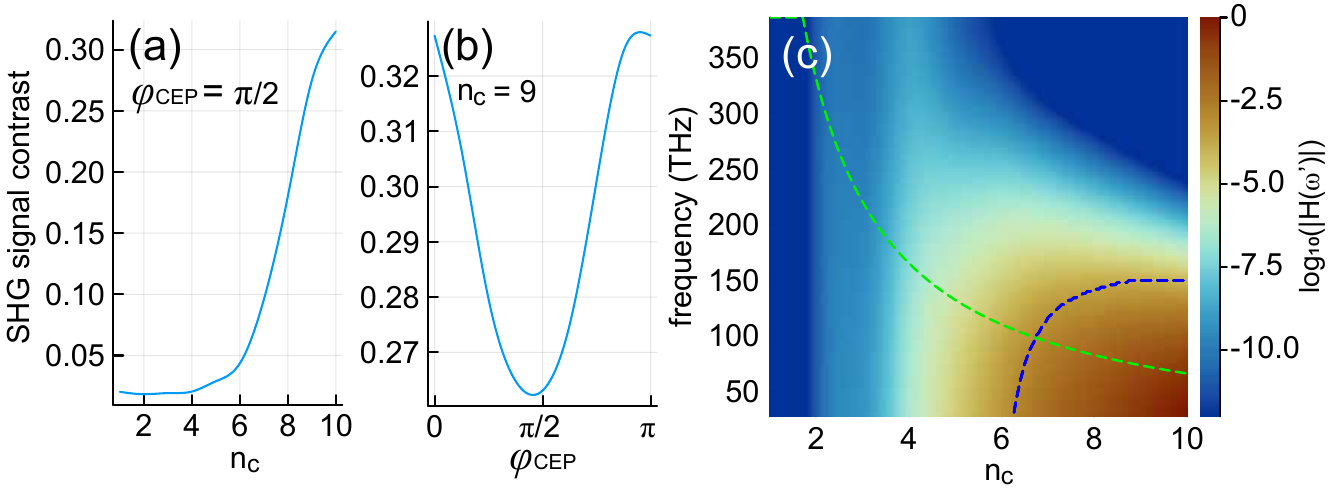}
    \caption{\label{fig:5} The signal contrast $C$ as a function of the pulse length $n_c$ for $\varphi_\mathrm{CEP}=\pi/2$ (a) and as a function of $\varphi_\mathrm{CEP}$ for $n_c=9$ (b). (c) The frequency response $|H(\omega^\prime)|$ \textit{vs} the pulse length $n_c$ at $\varphi_\mathrm{CEP}=\pi/2$, in units of dB. The maximum is normalized to 0 dB. The blue dashed line denotes the cutoff frequency $\omega^\prime_c$ where $|H(\omega^\prime)|$ declines to its half, and the green dashed line represents the traditional cutoff frequency given by the Fourier transform limit of the probe pulse (see in \cite{Supp}).
    }
\end{figure}

Alternatively, the probe pulse itself can break the underlying symmetry if the pulse duration is sufficiently short or the carrier-envelope phase is shifted. Such conditions preclude the exact cancellation of dipole radiation from adjacent half-cycles, manifesting as an intrinsic asymmetry in the sampling process. To provide a quantitative description, we return to the summation of all trajectories in Eq.(\ref{e4}), from which a compact analytical expression for the delay-dependent SHG amplitude can be derived \cite{Supp}:
\begin{equation}
        \begin{aligned}
            \mathcal{G}_{x}(2\omega_0; \tau) &= \mathcal{G}_{x}^{(0)}(2\omega_0) + \mathcal{G}_{x}^{(1)}(2\omega_0; \tau) \\ 
            &+ \mathcal{G}^{(2)}_{x}(2\omega_0; \tau) + \mathcal{G}^{(3)}_{x}(2\omega_0; \tau) + \cdots.
        \end{aligned}
    \label{e6}
\end{equation}
Here the zeroth-order contribution $\mathcal{G}_{x}^{(0)}(2\omega_0)$ represents the SHG background arising from the intrinsic asymmetry induced by the probe pulse itself, while the first-order term $\mathcal{G}_{x}^{(1)}(2\omega_0; \tau)=\int \mathrm{d} t \; E_\mathrm{THz}(t) h(\tau-t)$ is the linear response of the THz field with $h(t)$ being a convolution kernel related to the ionization gate. Therefore, we can define the SHG signal contrast $C=[\mathcal{G}_{x}(2\omega_0; \tau)-\mathcal{G}_{x}^{(0)}(2\omega_0)] / \mathcal{G}_{x}^{(0)}(2\omega_0)$. 
It exhibits a sharp decline with reduced pulse duration [Fig.5(a)] and oscillates with the carrier–envelope phase ($\varphi_\mathrm{CEP}$) [Fig.\ref{fig:5}(b)]. These results validate the preceding analysis.
To characterize the detection bandwidth, Fig.\ref{fig:5}(c) illustrates the frequency response $H(\omega^\prime) = \mathscr{F}\{\mathcal{G}_{x}(2\omega_0; \tau) / E_{\mathrm{THz}}(\tau)\}$ with varying pulse cycles $n_c$. The detection bandwidth limit is defined by the cutoff frequency $\omega^\prime_c$, where $|H(\omega^\prime)|$ falls to its half maximum (blue dashed line). 
Notably, when $n_c$ is sufficiently large, the SHG method demonstrates a significant bandwidth enhancement compared to the traditional Fourier limit of the probe pulse (green dashed line) \cite{Supp}. This underlies the physical mechanism enabling ultrabroadband THz detection.


In summary, we have elucidated the microscopic origin of the symmetry breaking responsible for SHG in the strong-field ionization regime. Specifically, the perturbation of the ionization rate by the target field induces an asymmetry in the emitted dipole radiation which hinders the destructive interference, and enables the emergence of the SHG signal. Our detailed analysis of this mechanism not only provides a physical rationale for the bandwidth enhancement observed in SHG-based THz-TDS, but also points out its limitations stemming from the intrinsic asymmetry induced by the probe field and from the back-reaction of the produced SHG. Our results show how one can optimize temporal resolution and signal contrast by steering the ionization gate in laser-matter interactions. Ultimately, exploiting these symmetry-breaking dynamics offers new degrees of freedom for coherent control, significantly broadening the scope and performance of field sampling in ultrafast terahertz and optical applications.


MK and QZL wish to thank Dr. K. Z. Hatsagortsyan for helpful discussions.
This work was supported by the Science Challenge Project (Grant No. TZ2025012) and by the National Natural Science Foundation of China (Grant Nos. 12334011, 12174284, 12541505 and 12475251).

\bibliography{sample.bib}

\end{document}